\documentclass[twocolumn,aps,tightenlines,floatfix,showpacs]{revtex4}

\usepackage{graphics}
\usepackage{bm}
\usepackage{amsmath}
\usepackage{amssymb}

\newcommand{\sfrac}[2]
{\ensuremath{\textstyle\frac{#1}{#2}}}

\begin{document}

\title{Comparison between two methods of solution of coupled equations
for low-energy scattering}

\author{K. Amos$^{(1)}$}
\email{amos@physics.unimelb.edu.au}
\author{S. Karataglidis$^{(1)}$}
\email{kara@physics.unimelb.edu.au}
\author{D. van der Knijff$^{(2)}$}
\email{dirk@unimelb.edu.au}
\author{L. Canton$^{(3)}$}
\email{luciano.canton@pd.infn.it}
\author{G. Pisent$^{(3)}$}
\email{gualtiero.pisent@pd.infn.it}
\author{J. P. Svenne$^{(4)}$}
\email{svenne@physics.umanitoba.ca}

\affiliation{$^{(1)}$ School of Physics, The University of Melbourne, 
   Victoria 3010, Australia}
\affiliation{$^{(2)}$ Advanced  Research   Computing,  Information
   Division, University of Melbourne, Victoria 3010, Australia}
\affiliation{$^{(3)}$ Istituto  Nazionale  di  Fisica  Nucleare,
   Sezione  di Padova, e\\  Dipartimento di Fisica  dell'Universit\`a
   di Padova, via Marzolo 8, Padova I-35131, Italia,}
\affiliation{$^{(4)}$ Department  of  Physics  and Astronomy,
   University  of Manitoba, and Winnipeg  Institute for   Theoretical 
   Physics, Winnipeg, Manitoba, Canada R3T 2N2}

\date{\today}

\begin{abstract}
  Cross sections from low-energy neutron-nucleus scattering have been
evaluated using a coupled channel theory of scattering.        Both a
coordinate-space    and  a    momentum-space    formalism   of   that 
coupled-channel theory are considered.      A simple rotational model 
of the channel interaction potentials is used to find results   using 
two relevant codes, ECIS97 and MCAS, so that they  may be   compared. 
The very same model is then used in the MCAS approach to quantify the 
changes that occur when  allowance is made for effects of the   Pauli 
principle.
\end{abstract}

\pacs{24.10-i;25.40.Dn;25.40.Ny;28.20.Cz}

\maketitle

\section{Introduction}
   In analyses of low-energy scattering data and in forming evaluated
nuclear data files, much use has been made of  programs  designed  to
solve equations  of coupled channels scattering theory. Programs such 
as CHUCK~\cite{Ku04} and ECIS~\cite{Ra88,Ra94}    seek such solutions
using a coordinate space representation of the scattering.   Versions
of  ECIS in fact are embedded within, or used with, such  large scale
analysis  programs  as GNASH~\cite{Yo92},  EMPIRE-II~\cite{He03}  and
TALYS~\cite{Ko04}, providing  basic input for the diverse evaluations
they make.  These codes, the ECIS codes in particular, use collective
model prescriptions for  the coupling interactions with   deformation
taken to second order for some cases. 

  It has long been known~\cite{Ma69,Gr96} that using these collective
model prescriptions violate the Pauli principle, and it has also been
argued that such violations could not be avoided.     However, it was
shown recently~\cite{Ca05} how the Pauli principle could be satisfied
with  a method of solution of the  coupled-channels  problem built in
momentum   space   using   separable   expansions   of   the coupling 
interactions.          That multi-channel algebraic scattering (MCAS) 
theory~\cite{Am03} when formed using Sturmians  that  are  orthogonal
to any Pauli blocked state as the expansion basis,     gave excellent
results for both the scattering  cross sections   and   sub-threshold
spectra for the examples considered:          protons and neutrons on
${}^{12}$C.            To create the appropriate set of Sturmians, an
orthogonalizing  pseudo-potential  (OPP) method was used~\cite{Am03}.
Violation of the Pauli  principle was shown to have serious effect on
results.    That raises concern about the application of interactions
and wave functions  generated by neglecting the Pauli principle  when 
interactions have been adjusted to give fits to low-energy scattering
data.  

   Herein, we report on a comparative study of the results of using a
coordinate space  program (ECIS97) and MCAS  (with and without taking
into account the Pauli principle) to see if   a) the calculations are
the same   when one seeks to perform the exact same evaluation   with
each,    b) for a typical low energy problem, how the Pauli principle
influences  the  results,  and    c) what  underlying structure of the 
compound system is inferred.

To compare the results of the two codes, we have used a simple (test)
model for the neutron-${}^{12}$C system.        We allow three target
states to define the  coupled  channels  in both the coordinate space
(ECIS97) and the momentum  space  (MCAS) evaluations.    They are the
ground ($0^+$), first excited state ($2^+$; 4.43 MeV), and the second
excited state ($0^+_2$; 7.67 MeV).   We also assume that the coupling
is effected by a  simple  rotational   model  scheme  having  only  a 
quadrupole deformation with   $\beta_2 = -0.52$ upon  a  purely  real
spherical Woods-Saxon potential~\cite{Ra94} given in MeV.  All length
parameters are expressed in fm, and the deformed field form is
\begin{eqnarray}
V(r) &=& -49.92\, f(r)\, +\, \left(\frac{\hbar}{m_\pi c}\right)^2
6\, {\mathbf \sigma \cdot \nabla f(r) \times \frac{1}{i}\nabla}
\,\,\,\,\,\,\,\,\,\, 
\label{Equation1}
\\
f(r) &=& \left[1 + \exp{\left(\frac{r - 2.885}{0.63}\right)}
\right]^{-1}.
\end{eqnarray}
In   the   MCAS   evaluation   the spin-orbit term is reduced to  the 
{\bf l$\cdot$s} form.   This potential is fixed  for all calculations
that we have made and whose results are reported herein.

  For the comparative study of $n+^{12}$C in a rotational model, MCAS
carries the deformation up to second order.     ECIS~\cite{Ra88,Ra94}
allows deformation to second order with various vibration       model 
specifications of the channel interactions,     but with the rotation 
model, the expansion of the nuclear deformation is only taken to first   
order.

\section{General remarks about ECIS and MCAS}

      Details of what these two codes calculate are  presented in the
literature and so only brief comment is given here. For ECIS we refer
the reader to the documentation~\cite{Ra88,Ra94} for a more  detailed 
description. With MCAS there are three publications to consider.  The 
first~\cite{Am03} gives a detailed description of the method  and the  
model interactions chosen for the application made.               The 
second~\cite{Ca05} highlights how that process corrects  a collective 
model prescription of the scattering to allow for the Pauli principle, 
and   therein it is  shown   just how crucial that is if a physically  
significant interaction is to be defined. Finally, in Ref.~\cite{Pi05} 
the physics that can be extracted by using 
the MCAS scheme is highlighted,  but  only  when the  Pauli principle 
effects are treated. 

      ECIS97 has been constructed to use a wide range of (collective)
model structures  to describe   the   nuclear interaction matrices of
potentials, $V_{c'c}(r)$.      MCAS on the other hand is still in its
infancy and to date   the  only  working program is one that inputs a
rotational (collective) model matrix of potentials.    Development to
incorporate a vibration model for the target spectrum as well as   to
use  shell  model wave functions to define  the  matrices of coupling 
potentials is proceeding.       However, to make a comparison between
two codes we consider only the case of a simple rotation model scheme.

We specify  the complete channel index by 
\mbox{$c:(\ell(\sfrac{1}{2})j,I;J^\pi)$}, 
which couples the incident partial
wave  angular momenta $\{\ell(\frac{1}{2})j\}$ to the target spin $I$ to
get the total system spin-parity $J^\pi$.  The last is conserved in
the scattering process.   ECIS97 solves the  coupled-channels problem 
in coordinate space so that the defining equations  have   the   form  
(Eq.~(17) in Ref.~\cite{Ra88}),
\begin{equation}
 \frac{\hbar^2}{2\mu}
\left[\frac{d^2}{dr^2} - \frac{\ell(\ell+1)}{r^2} + k^2\right] f_c(r)
= \sum_{c'} V_{c,c'}(r)\, f_{c'}(r)\ ,
\end{equation}
where the notation is as usual. As noted~\cite{Ra88},       the wave 
functions $f_c(r)$ have asymptotic forms for   large $r$ (Eq.~(8) in 
Ref.~\cite{Ra88})
\begin{equation}
f_{l,j}(r) = F_l(\eta, kr) 
+ C_{l,j} \left[G_l(\eta, kr) + i F_l(\eta, kr) 
\right]\ ,
\end{equation}
where, with   $\eta$ being the Sommerfeld parameter, $F_l,\, G_l$ are
the regular and irregular (at the origin) Coulomb functions.      The
solutions in the case of closed channels are the appropriate decaying
forms.

 With the rotational model for the matrices of potentials, in  ECIS97
a first order  multipole expansion is considered, namely
\begin{equation}
R = R_0 
\left[1 + \sum_\lambda \beta_\lambda Y_{\lambda, 0}^\star(\Omega_A) 
\right] \, ,
\end{equation}
so  that  the  operator form of the  projectile-nucleus interactions 
becomes
\begin{equation}
V(r, \Omega_A) = V_0(r) + \sum_{\lambda} V_\lambda(r) 
{\mathbf Y}_{\lambda}^\star(\Omega_A) \cdot 
{\mathbf Y}_{\lambda}(\Omega_r)\ . 
\end{equation}
     When only the quadrupole moment defines the test model, an 
ECIS97 run should then coincide with an   MCAS   calculation in which
deformation is limited to first order.

  The details of the MCAS approach are published in Ref.~\cite{Am03},
and so we do not repeat them here.           It suffices to note that
deformation of the interaction from the rotation model    is taken to
second   order  and  the  program  allows  flexibility  in the forms; 
permitting   parity,  orbital  angular  momentum,   and  target  spin 
dependences.  For the test  model  calculations, the results of
which are reported herein, such flexibility has not been exploited.

\section{Results and discussion}

\begin{figure}[t]
\scalebox{0.5}{\includegraphics{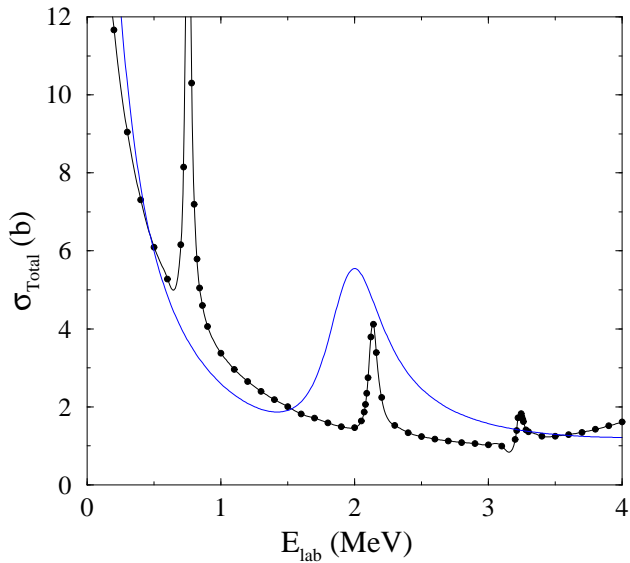}}
\caption{\label{Fig1}(Color online)
The results from using ECIS97 to evaluate the test model  cross
sections for the $n+{}^{12}$C system. 
The filled circles connected by a line is the cross section
found from the ECIS coupled channels calculations while the solid curve
is the result when coupling is set to zero (the ground state potential
scattering calculation).}
\end{figure}

    The program ECIS97 was run for the test model at a series of 
(laboratory) energies $E_{lab}$ from 0.1 MeV to 4.0 MeV.   
The results are displayed  in Fig.~\ref{Fig1} by the filled 
circles connected by a (spline) curve and reveal
three resonances near 0.7, 2.1, and 3.3 MeV.  
The solid curve in that figure is the cross section found  
from ECIS97 calculations made using 
the same spherical potential but considering only the elastic
channel. That is the basic optical model result in which there is a
shape, or single particle, resonance 
centered  about $E_{\rm lab} = 2$ MeV.       Clearly the inclusion of
channel coupling changes these cross sections significantly. So  
the results we compare next are ones of a significant coupled channel
problem and not ones that might be obtained simply by adjustments of
the parametric form of the ground state (optical) potential.

ECIS calculations usually are made with the full Thomas form of the 
spin-orbit interaction. However in the write-up of that code~\cite{Ra88}
it is shown how one can limit calculations so that the 
${\bf l\cdot s}$ form is used. That form is what we have
incorporated (so far) in MCAS. We have made ECIS calculations both
with the full Thomas and with the ${\bf l \cdot s}$ forms.
The two calculated cross  sections are in very 
good agreement for most of the energy  range  and  only   the  strong
low-energy  $\frac{5}{2}^+$ resonance is slightly shifted in its 
centroid by the reduction to the simplest spin-orbit form.        
These results  corroborate findings in 
previous studies~\cite{Sh68}  that only at higher energies,  and  for 
observables directly linked to   inelastic-channel interactions, does 
use    of    the    full    Thomas   term  rather than the 
$\mathbf{l \cdot s}$ form have some effect.   Even then, those effects 
are very small and  essentially  with  the  forward  angle  spin 
dependent observables, such as the analyzing powers.

\begin{figure}[t]
\scalebox{0.5}{\includegraphics{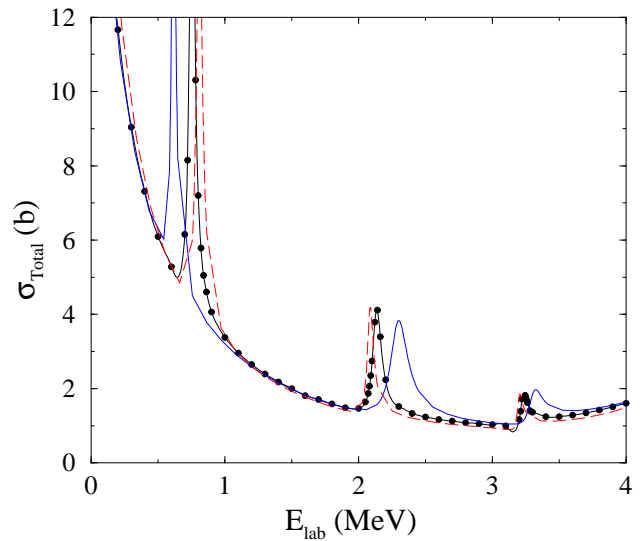}}
\caption{\label{Fig2}(Color online)
The $n+{}^{12}$C cross section results from using MCAS theory to first 
order in deformation   (dashed curve) and for deformation taken 
to second order (solid curve),     compared 
with those found by using the ECIS97 program (filled circles
connected by a line).}
\end{figure}

    First  MCAS calculations have been made using the same test model, 
and the fixed interaction given in Eq.~(\ref{Equation1}) but with the
(${\bf l \cdot s}$) form for the spin-orbit components
and   without   accounting for the Pauli principle. 
In Fig.~\ref{Fig2},  these 
results are compared with those found using the ECIS97 code. The ECIS
results again are displayed by the filled circles connected by a
solid line and  there are two MCAS results. The first, displayed by the 
solid
curve, involved deformation taken through second   order~\cite{Am03}.  
It agrees with the background found from the ECIS calculation and also 
has the same three resonances though 
their energy centroids are shifted. 
The second 
MCAS result, depicted by the dashed curve,   was obtained by limiting 
deformation to first order. This result is in better agreement with the  
ECIS cross section,     both background and resonances (centroids and 
widths).
Slight differences must be allowed since the two codes involve  quite
different numerics and associated accuracies.  
We consider 
the whole
set of results to be close enough to claim that the   two  codes  are
equivalent in what they evaluate.

Two conclusions may be drawn from the results found so far. First, when 
the test  model is used in exactly the   same  way  in  finding 
solutions of the coupled-channel problem using the coordinate  space
approach~\cite{Ra88}     and     the     momentum space approach with 
MCAS~\cite{Am03}, the scattering cross sections agree very well.  The
smooth   background  as  well  as  the  specific  resonances that can
be generated with ECIS are found with the MCAS run.      The second
conclusion evident from comparison of the two   MCAS results is that,
with deformation of $\beta_2=-0.52$ which is realistic for the actual
system, a first order approach is insufficient.

\begin{figure}[t]
\scalebox{0.5}{\includegraphics{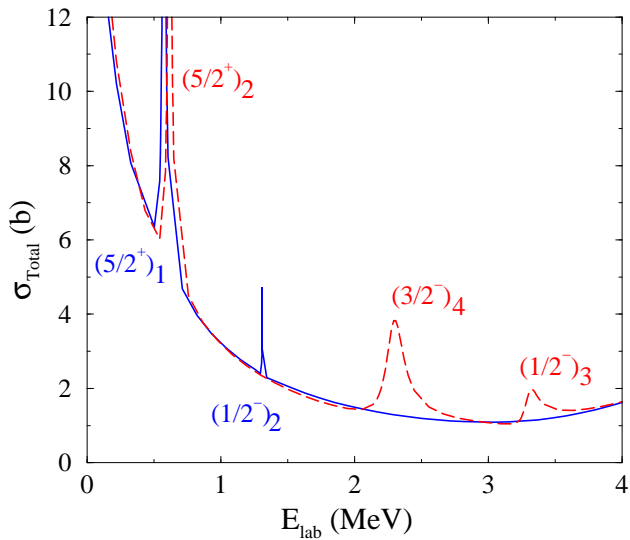}}
\caption{\label{Fig3}(Color online)
The $n+{}^{12}$C cross section results from using    MCAS theory with
(solid curve)   and   without  (dashed curve) using the OPP method to
prevent violation of the Pauli principle.}  
\end{figure}

We have shown that the two programs evaluate equivalent cross sections
but those evaluations are equivalently in error as the effects of the
Pauli principle~\cite{Ca05} have been ignored.  We now consider just how
important it is to include the Pauli principle
and how the associated blocking mechanism works with this test model.
As noted earlier,
in the MCAS approach,
an OPP method can be used to ensure that there is no violation of the
Pauli principle.  The OPP method ensures that the  Sturmians  used an
expansion set in the  MCAS approach   are orthogonal to all states in
which   the  incoming  nucleon  would  be trapped into an orbit fully
occupied by nucleons in the target. Using such a conditioned Sturmian
function set to solve the MCAS theory of  coupled equations gives the
cross section displayed by the solid curve in Fig.~\ref{Fig3}.   That
is compared with the MCAS result shown previously and found   without
using the OPP and taking deformation also to second order.       That 
latter result is portrayed by the dashed curve in    Fig.~\ref{Fig3}.
The changes seen are dramatic. 

To discuss them, first it is important
to note that the MCAS theory~\cite{Am03} embodies a resonance finding 
scheme with which all subthreshold and resonance states, no matter how
narrow any of the latter may be, that lie within any energy range 
selected for study will be found and their spin-parities, energy 
centroids, and widths determined. Furthermore the order number of each
can be obtained. The order number ($r$) identifies that 
there are $r-1$ bound states/resonances of that given $J^\pi$ lying 
below in the spectrum of the compound system.
In Fig.~\ref{Fig3} then, each resonance is identified by its value of 
$\left(J^\pi\right)_r$. While the background cross section calculated
with and without Pauli blocking is essentially unchanged, the resonance 
properties are drastically altered. Both calculations give a 
$\frac{5}{2}^+$ resonance near 0.6 MeV but the number order  differs.
Then, the $\frac{3}{2}^-$ resonance disappears while the
$\frac{1}{2}^-$ resonance relocates to lower energy, changes its 
order number to 2, and has a much narrower width when the effects
of the Pauli principle are considered. The prime effect of including 
the Pauli principle is to remove numerous spurious states from the 
spectrum. However, it also changes the underlying structure of what
states remain~\cite{Pi05}. In that reference, the tracking of states and
resonances as deformation is decreased to zero revealed the basic origin
of each state and resonance. With the test model, set so that a direct
comparison between two methods of solving coupled channels problems
can be made, we show in Table~\ref{Zbt}, the full spectra that have 
been obtained using MCAS with and without the OPP and in the zero deformation
limit. This table is similar 
to that given previously~\cite{Pi05} and which was found with a matrix
of interaction potentials that gave an excellent fit to data. However,
it is important to present these values, not only as they are specifically 
those from the test model we have used, but also as some of the values bear upon
conclusions to be drawn from the results shown in Fig.~\ref{Fig3}.
\begin{table}[ht]
\caption{\label{Zbt} 
The spectra found with MCAS when $\beta_2 \longrightarrow 0$.  
In the first column, the numerical labels for the spurious states
are presented in the brackets, \{n\}. The arrows in the second column
indicate the Pauli-allowed states obtained when the OPP is applied.
The subscript $r$ is the order number of each state and resonance.} 
\squeezetable
\begin{ruledtabular}
{\small
\begin{tabular}{ccrc}
label & $\left(J\right)^\pi_r$ & Energy$\;\;$ 
& $^{12}{\rm C} + (n\ell_j)$\\
\hline
\{1\} &$\left(\frac{1}{2}\right)^+_1$ & -23.50$\;\;$ 
& $0^+_1 + 0s_{\frac{1}{2}}$\\
\{2\} &$\left(\frac{3}{2}\right)^+_1,\left(\frac{5}{2}\right)^+_1$ 
& -19.07$\;\;$
& $2^+_1 + 0s_{\frac{1}{2}}$\\
\{3\} &$\left(\frac{1}{2}\right)^+_2$ & -15.85$\;\;$
& $0^+_2 + 0s_{\frac{1}{2}}$\\ 
\{4\} &$\left(\frac{3}{2}\right)^-_1$ &  -9.73$\;\;$
& $0^+_1 + 0p_{\frac{3}{2}}$\\
5 &$\left(\frac{1}{2}\right)^-_1\rightarrow
\left(\frac{1}{2}\right)^-_1 $ & -5.92$\;\;$
& $0^+_1 + 0p_{\frac{1}{2}}$\\
\{6\} &$\left(\frac{1}{2}\right)^-_2, \left(\frac{3}{2}\right)^-_2, 
\left(\frac{5}{2}\right)^-_1, \left(\frac{7}{2}\right)^-_1$ 
& -5.29$\;\;$
& $2^+_1 + 0p_{\frac{3}{2}}$\\
\{7\} &$\left(\frac{3}{2}\right)^-_3$ & -2.07$\;\;$
& $0^+_2 + 0p_{\frac{3}{2}}$\\
8 &$\left(\frac{3}{2}\right)^-_4, \left(\frac{5}{2}\right)^-_2
\rightarrow \left(\frac{3}{2}\right)^-_1, \left(\frac{5}{2}
\right)^-_1$ & -1.48$\;\;$
& $2^+_1 + 0p_{\frac{1}{2}}$\\ 
9 &$\left(\frac{1}{2}\right)^-_3\rightarrow 
\left(\frac{1}{2}\right)^-_2$ & 1.74$\;\;$
& $0^+_2 + 0p_{\frac{1}{2}}$\\ 
10&$\left(\frac{5}{2}\right)^+_2\rightarrow 
\left(\frac{5}{2}\right)^+_1$ & 2.08$\;\;$
& $0^+_1 + 0d_{\frac{5}{2}}$\\
\end{tabular}
}
\end{ruledtabular}
\end{table}

For simplicity of discussion each state or group of states at a
given value are identified by  a label number in the first    column.
The states associated with labels set in curly brackets arise   from
Pauli violation and are numerically removed by the OPP method. In the
second column the arrow indicates the  Pauli-allowed states,  all but 
the lowest of which are reduced in order number due to Pauli blocking. 
The energy gaps between, and spin-parities of,   these states lead to
the base prescription    given     in  the column on the far right of
Table~\ref{Zbt}.  The energy gaps in the zero
deformation limit relate directly to the target spectrum values   and
the single nucleon state binding energies. Of relevance in this 
discussion is that the $\left(\frac{1}{2}^-\right)_2$ state in group
labelled \{6\} is spurious.  The allowed state 
$\left(\frac{1}{2}^-\right)_3$ of the set (the entry in group 9 in 
Table~\ref{Zbt}) then becomes the $\left(\frac{1}{2}^-\right)_2$ state 
after application of the OPP method.  Hence there is the reduction in 
order number of
the calculated resonance state of that spin-parity shown in 
Fig.~\ref{Fig3}. Moreover, and associated with the removal of a basic 
spurious state of that spin-parity, with finite deformation forming 
admixtures to yield the end result, there will be no spurious component 
then in the resultant narrow resonance centered near 1.3 MeV. The
change in character of that resonance due to the Pauli principle is
evident. Likewise the lowest three $\frac{3}{2}^-$ states also are
Pauli forbidden so the remaining allowed state is one of the subthreshold
compound nuclear states and there is no resonance of that spin-parity
in the resultant cross section in contradiction to the result found
without taking the Pauli principle into account.  Finally there is one 
spurious $\frac{5}{2}^+$ state in the spectrum that has been removed
and as the remaining state of that spin-parity is basically built as
the $0d_{\frac{5}{2}}$ neutron coupled to the ground state of ${}^{12}$C,
there is no great change in centroid energy when the Pauli principle is
considered.
So, there are many spurious states when the Pauli principle
is violated.         Worse, there are spurious states having the same
spin-parities as those to be found when the    Pauli    principle  is
preserved  in the calculations.  With deformation coupling, these
basis states mix to determine  that to be deemed the physical result.

If either code (used without Pauli correction) found that the
simple interaction actually gave fits to cross-section data, then
that interaction and, more importantly,   the relative wave functions
derived from it, would be wrong.     One would need to invoke the OPP
approach (or an equivalent) and then make a further  parameter search
to find an interaction that leads to a fit to the data.     But it is
important to note that the background cross section itself   does not
provide selectivity as it is dominated by $s$-wave scattering.

\section{Conclusions}

        In this paper we have compared the results when two different
coupled-channel approaches (MCAS and ECIS) are used.  To achieve that
we have considered a simple potential for the description of  the 
nucleon-$^{12}$C dynamics which includes low-lying excitations of the 
target  in  terms  of  a  collective, rotational-type model where the 
quadrupole deformation $\beta_2$ has been set to  a (realistic) value 
of $-0.52$. Within the unavoidable small differences that remain in 
the construction of the programs, for this particular  case  we  have 
shown 
that the results of the two approaches are essentially equivalent.

 However, with the MCAS approach we could include also the effects of 
second-order contributions in the deformation parameter. They lead to
substantial changes to the cross section. Even more importantly, with
the MCAS approach we could eliminate the spurious  states that appear 
if one ignores the effects of the    Pauli    principle    with   the 
Schr{\"o}dinger equation. In the MCAS method one can take account  of  
the effects due to the identity between the projectile nucleon    and 
the nucleons in the target by applying a suitable generalization   of
the orthogonalizing pseudo-potential method. The effects due to Pauli
principle are very significant; greatly influencing the overall 
structure of
the cross-section and  changing  completely the  resonant  and  bound
spectra of the compound system  associated  with a fixed interaction.

A distinctive feature of the MCAS approach is that, by 
study of the spectra as $\beta_2\rightarrow 0$, it allows the unphysical
nature of the spurious states to be illustrated.  Doing that in a 
previous study~\cite{Pi05}  emphasized
the need for their elimination from the coupled-channel dynamics. 

     Another interesting feature of the MCAS approach is that one can 
systematically   track   all   resonances  and bound-state structures 
contained in the compound system.        This feature is particularly 
welcome   for   the  specific problem we  have considered and is a 
consequence of the use  of  Sturmian  states in the expansion scheme.
Even  the  most  narrow  resonant scattering state can be numerically 
determined, its spin-parity and width can be easily evaluated without
the need to organize an extremely fine (and extremely time consuming)
energy spanning of the   $S$-matrix to seek rapid increases in  phase
shifts.    Closely related to this property is the capacity to assess
the order number of a given resonance which indicates      how many other
resonances and bound states with the same spin and parity lie   below
the one considered. This   parameter
is   important    within   the  process of data evaluation, since 
fitting 
procedures in coupled-channel   calculations that ignore the need for 
dealing with an entire ensemble of physical states  (without spurious
entries) have very little physical relevance.

     In summary, the MCAS approach, albeit still in its infancy, is a 
promising means to study low energy nuclear reaction  cross  sections
since it allows treatment of the Pauli principle in a simple  manner,
it facilitates solution of sub-threshold spectra as well as  defining
resonance behavior due to coupled-channel effects,  and encompasses a
procedure which finds all resonances produced in the selected  
energy interval.   On the other hand,   the coordinate space
coupled-channels programs currently in use need upgrading at least to
incorporate effects of the Pauli principle before their  interactions
and associated relative motion wave functions may be used        with
confidence of physical significance.     Whether some scheme, such as 
supersymmetric quantum mechanics,        can be found to effect that 
upgrade is a major problem for developers and users of those codes.

\begin{acknowledgments}
  This research was supported by a grant from the Australian Research
Council, by a merit award with the Australian Partners for   Advanced
Computing, by the Italian MIUR-PRIN Project      ``Fisica Teorica del
Nucleo e dei Sistemi a Pi\`u Corpi'', and by the Natural Sciences and
Engineering Research Council (NSERC), Canada. 
\end{acknowledgments}

\bibliography{Ecis-CCCR-5}
\end{document}